# Applications of Random Parameter Matrices Kalman Filtering in Uncertain Observation and Multi-Model Systems

Dandan Luo    Yunmin Zhu*

Department of Mathematics

Sichuan University

Chengdu, Sichuan 610064, China

rodan531@163.com    ymzhu@scu.edu.cn

**Abstract**: This paper considers the Linear Minimum Variance recursive state estimation for the linear discrete time dynamic system with random state transition and measurement matrices, i.e., random parameter matrices Kalman filtering. It is shown that such system can be converted to a linear dynamic system with deterministic parameter matrices but state-dependent process and measurement noises. It is proved that under mild conditions, the recursive state estimation of this system is still of the form of a modified Kalman filtering. More importantly, this result can be applied to Kalman filtering with uncertain observations as well as randomly variant dynamic systems with multiple models.

## 1 Introduction

Linear discrete time system with random state transition and observation matrices arise in many areas such as radar control, missile track estimation, satellite navigation, digital control of chemical processes, economic systems. Koning [1] gave the Linear Minimum Variance recursive estimation formulae for the linear discrete time dynamic system with random state transition and measurement matrices without rigorous derivation. Such system can be converted to a linear dynamic system with deterministic parameter matrices and state-dependent process and measurement noises. Therefore, the conditions of standard Kalman Filtering are violated and the recursive formulae (for example, in

*** Corresponding author**. Supported in part by NSF of China (#60374025 and 60328306) and SRFDP (#20030610018).

[1]) can not be derived directly from the Kalman Filtering Theory. A rigorous analysis in this paper shows that under mild conditions, the converted system still satisfies the conditions of standard Kalman Filtering; therefore, the recursive state estimation of this system is still of the form of a modified Kalman filtering.

More importantly, this result can be applied to many practical problems related to Kalman filtering. Recently, the Kalman filtering with uncertain observation attracted extensive attentions [4, 5, 6]. There are two types of uncertain observations in practice. The first one is that the estimator can exactly know whether the observation fully or partially contains the signal to be estimated, or just contains noise alone (for example, see [4]). By directly using the optimal estimation theory, the Kalman filter for the first type of uncertain observations can be derived easily. The other uncertain observations belong to the second type, i.e., the estimator cannot know whether the observation fully or partially contains the signal to be estimated, or just contains noise alone, but the occurrence probabilities of each case is known. By applying the random measurement matrix Kalman filtering, we can derive the Kalman filter with the second type of uncertain observations, which is more general than that in [4, 6].

On the other hand, in practical applications, people may face a Multi-Model(**MM**) Dynamic Process very often. The **MM** Dynamic Process is best described in terms of randomly variant dynamic systems. Such a system is one that can be suitably described in a hybrid space $R^{n_x} \times \mathcal{S}$, the Cartesian product of the continuous-valued base state space $R^{n_x}$ and a discrete finite set $\mathcal{S}$, the collection of the finite system modes which characterize the behavior patterns of the system. A randomly variant dynamic system thus distinguishes itself from conventional systems in its imbedded random jump process which governs the random transition of its system behavior patterns. Many real-world problems can be successfully formulated in terms of such systems. Typical examples can be found in systems subject to piecewise linearization of nonlinear systems, maneuvering target tracking, reconfigurable systems, etc. The **MM** dynamic process were considered by many researchers (for example, see [7]-[11]). Although the possible models there are quite general and possibly depend on the state, only suboptimal algorithms were proposed in the past a few decades. However, some of the **MM** systems, although they are not dependent on the state and somewhat restrictive than the the models considered in [8, 10], can be reduced to the dynamic models with random transition matrix, therefore, the optimal filter can be given directly according to the random transition matrix Kalman filtering given here. The simulation results support the analysis in this paper.

The remainder of this paper is organized as follows. In Section 2, we present the random parameter matrices Kalman filtering. In Section 3, we formalize a general model of Kalman filtering with uncertain observations, and derive the optimum linear recursive estimators by applying the random Kalman filtering. And two application examples are provided to give an intuitive understanding of the results. In Section 4, we formalize the multiple-model dynamic process as a process with random transition matrix and provide the optimal real-time estimator for this case. In Section 5, simulation examples are given for the models given in Section 3 and Section 4. Finally, in Section 6, we present our conclusions.

## 2 Random Parameter Matrices Kalman Filtering

Consider a discrete time dynamic system

$$\mathbf{x}_{k+1} = F_k \mathbf{x}_k + \nu_k, \tag{1}$$

$$\mathbf{y}_k = H_k \mathbf{x}_k + \omega_k , \ k = 0, 1, 2, \cdots, \tag{2}$$

where $\mathbf{x}_k \in \mathcal{R}^r$ is the system state, $\mathbf{y}_k \in \mathcal{R}^N$ is the measurement, $\nu_k \in \mathcal{R}^r$ is the process noise, and $\omega_k \in \mathcal{R}^N$ is the measurement noise. The subscript $k$ is the time index. $F_k \in \mathcal{R}^{r \times r}$ and $H_k \in \mathcal{R}^{N \times r}$ are random matrices.

We assume the system has the following statistical properties: $\{F_k, H_k, \nu_k, \omega_k, k = 0, 1, 2, \cdots\}$ are all sequences of independent random variables temporally and across sequences as well as independent of $\mathbf{x}_0$. Moreover, we assume $\mathbf{x}_k$ and $\{F_k, H_k, k = 0, 1, 2, \cdots\}$ are independent mutually. The initial state $\mathbf{x}_0$, the noises $\nu_k, \omega_k$, and the parameter matrices $F_k, H_k$ have the following means and covariances

$$E(\mathbf{x}_0) = \mu_0, \quad E(\mathbf{x}_0 - \mu_0)(\mathbf{x}_0 - \mu_0)^T = P_0, \tag{3}$$

$$E(\nu_k) = 0, \quad E(\nu_k \nu_k^T) = R_{\nu_k}, \quad E(\omega_k) = 0, \quad E(\omega_k \omega_k^T) = R_{\omega_k}, \tag{4}$$

$$E(F_k) = \bar{F}_k, \quad Cov(f_{ij}^k, f_{mn}^k) = C_{f_{ij}^k f_{mn}^k}, \tag{5}$$

$$E(H_k) = \bar{H}_k, \quad Cov(h_{ij}^k, h_{mn}^k) = C_{h_{ij}^k h_{mn}^k}, \tag{6}$$

where $f_{ij}^k$ and $h_{ij}^k$ are the $(i,j)th$ entries of matrices $F_k$ and $H_k$, respectively.

Rewrite $F_k$ and $H_k$ as

$$F_k = \bar{F}_k + \tilde{F}_k, \tag{7}$$

$$H_k = \bar{H}_k + \tilde{H}_k. \tag{8}$$

Substituting (7), (8) into (1), (2) converts the original system to

$$\mathbf{x}_{k+1} = \bar{F}_k \mathbf{x}_k + \tilde{\nu}_k, \tag{9}$$

$$\mathbf{y}_k = \bar{H}_k \mathbf{x}_k + \tilde{\omega}_k, \tag{10}$$

where

$$\begin{aligned} \tilde{\nu}_k &= \nu_k + \tilde{F}_k \mathbf{x}_k \\ \tilde{\omega}_k &= \omega_k + \tilde{H}_k \mathbf{x}_k \end{aligned} \tag{11}$$

System (9), (10) has deterministic parameter matrices, but the process noise and observation noise are dependent on the state; therefore, this would not satisfy the well-known assumptions of standard Kalman filtering.

In the following, we will derive the recursive state estimate of the new system, which is of the form of a modified Kalman filtering. We present two lemmas first, and leave their proofs in the appendix of the paper.

**Lemma 1.** Suppose random matrix F and random vector $\mathbf{x}$ are independent, then

$$E(F\mathbf{x}\mathbf{x}^T F^T) = E(FE(\mathbf{x}\mathbf{x}^T)F^T)$$

**Lemma 2.**

(a) $E(\tilde{\nu}_k) = 0$, $E(\tilde{\omega}_k) = 0$;

(b1) $E(\mathbf{x}_0 \tilde{\nu}_k^T) = 0$, (b2) $E(\mathbf{x}_0 \tilde{\omega}_k^T) = 0$;

(c1) $E(\tilde{\nu}_k \tilde{\nu}_l^T) = 0$, (c2) $E(\tilde{\omega}_k \tilde{\omega}_l^T) = 0$, (c3) $E(\tilde{\nu}_k \tilde{\omega}_l^T) = 0 \quad \forall k \neq l$;

(d) $E(\tilde{\nu}_k \tilde{\nu}_k^T) = R_{\tilde{\nu}_k}$, $E(\tilde{\omega}_k \tilde{\omega}_k^T) = R_{\tilde{\omega}_k}$,

where

$$R_{\tilde{\nu}_k} = R_{\nu_k} + E(\tilde{F}_k E(\mathbf{x}_k \mathbf{x}_k^T) \tilde{F}_k^T),$$
$$R_{\tilde{\omega}_k} = R_{\omega_k} + E(\tilde{H}_k E(\mathbf{x}_k \mathbf{x}_k^T) \tilde{H}_k^T).$$

By Lemma 2, system (9), (10) satisfies all conditions of the standard Kalman Filtering. Hence, we have the following theorem ([2, 3]) immediately.

**Theorem 1.** The Linear Minimum Variance recursive state estimation of system (9), (10) is given by

$$\begin{aligned}
\mathbf{x}_{k+1|k+1} &= \mathbf{x}_{k+1|k} + K_{k+1}(y_{k+1} - \bar{H}_{k+1}\mathbf{x}_{k+1|k}) \\
\mathbf{x}_{k+1|k} &= \bar{F}_k \mathbf{x}_{k|k} \\
K_{k+1} &= P_{k+1|k} \bar{H}_{k+1}^T (\bar{H}_{k+1} P_{k+1|k} \bar{H}_{k+1}^T + R_{\tilde{\omega}_k})^+ \\
P_{k+1|k} &= \bar{F}_k P_k \bar{F}_k^T + R_{\tilde{\nu}_k} \\
P_{k+1} &= (I - K_{k+1}\bar{H}_{k+1})P_{k+1|k} \\
R_{\tilde{\nu}_k} &= R_{\nu_k} + E(\tilde{F}_k E(\mathbf{x}_k \mathbf{x}_k^T) \tilde{F}_k^T) \\
R_{\tilde{\omega}_k} &= R_{\omega_k} + E(\tilde{H}_k E(\mathbf{x}_k \mathbf{x}_k^T) \tilde{H}_k^T)
\end{aligned}$$

$$E(\mathbf{x}_{k+1}\mathbf{x}_{k+1}^T) = \bar{F}_k E(\mathbf{x}_k \mathbf{x}_k^T) \bar{F}_k^T + E(\tilde{F}_k E(\mathbf{x}_k \mathbf{x}_k^T) \tilde{F}_k^T) + R_{\nu_k}$$

$$\mathbf{x}_{0|0} = E\mathbf{x}_0, \ P_0 = Var(\mathbf{x}_0), \ E(\mathbf{x}_0 \mathbf{x}_0^T) = E\mathbf{x}_0 E\mathbf{x}_0^T + P_0,$$

where the superscript " + " denotes the Moore–Penrose pseudo inverse.

**Remark 1.** Compared with the standard Kalman filtering, random parameter matrices Kalman filtering has one more recursion of $E(\mathbf{x}_{k+1}\mathbf{x}_{k+1}^T)$. By Theorem 1, we eventually have to compute $E(\tilde{F}_k E(\mathbf{x}_k \mathbf{x}_k^T) \tilde{F}_k^T)$ and $E(\tilde{H}_k E(\mathbf{x}_k \mathbf{x}_k^T) \tilde{H}_k^T)$ and their analytical expressions are given as:

$$E(\tilde{F}_k E(\mathbf{x}_k \mathbf{x}_k^T) \tilde{F}_k^T)_{(m,n)}$$
$$= \sum_{i=1}^r C_{f_{n1}^k f_{mi}^k} X_{i1}^k + \sum_{i=1}^r C_{f_{n2}^k f_{mi}^k} X_{i2}^k + \cdots + \sum_{i=1}^r C_{f_{nr}^k f_{mi}^k} X_{ir}^k, \quad m,n = 1,2,\cdots,r$$

$$E(\tilde{H}_k E(\mathbf{x}_k \mathbf{x}_k^T) \tilde{H}_k^T)_{(m,n)}$$
$$= \sum_{i=1}^r C_{h_{n1}^k h_{mi}^k} X_{i1}^k + \sum_{i=1}^r C_{h_{n2}^k h_{mi}^k} X_{i2}^k + \cdots + \sum_{i=1}^r C_{h_{nr}^k h_{mi}^k} X_{ir}^k, \quad m,n = 1,2,\cdots,N$$

where $X^k = E(\mathbf{x}_k \mathbf{x}_k^T)$.

## 3 Application to A General Uncertain Observation

Consider a system

$$\mathbf{x}_{k+1} = F_k \mathbf{x}_k + \nu_k \tag{12}$$

$$\mathbf{y}_k = \sum_{i=1}^l I_{\{\gamma(k)=i\}} H_k^i \mathbf{x}_k + \sum_{i=1}^l I_{\{\gamma(k)=i\}} \omega_k^i, \tag{13}$$

where all the parameter matrices are non-random and a set of multiple observation equations is selected to represent the possible observation case at each time. The random variable $\gamma_k$ is defined to formulate which measurement matrix is chosen at time $k$ and the value of $\gamma_k$ is either observable or unobservable. If $\gamma_k = i$, the measurement matrix is $H_k^i$ and the observation noise corresponds to $\omega_k^i$. When the value of $\gamma_k$ is observable at each time $k$, this is an uncertain observation of the first type and the state estimation with measurement equation (13) is converted to

$$\mathbf{y}_k = H_k^i \mathbf{x}_k + \omega_k^i, \tag{14}$$

which is obviously the classical Kalman filtering, i.e., the least mean square estimate using the various available observation of $\mathbf{y}_k$. To show the applications of the random measurement matrix Kalman filtering, we focus on the the second type of uncertain observations, i.e., in (13), $\gamma_k$ is unobservable at each time $k$, but the probability of the occurrence of every available measurement matrix is known.

Consider that in (13), $\gamma_k$ is unobservable at each time $k$, but the probability of the occurrence of each measurement matrix is known. Obviously, (2) is a more general form of (13) because only expectation and covariance of $H_k$ in (2) are known (see (6)) other than its distribution. The expectation of $H_k$ can be expressed as:

$$\bar{H}_k = \sum_{j=1}^{l} p_k^j H_k^j \tag{15}$$

$$\tilde{H}_k^i = H_k^i - \bar{H}_k, \text{ with probability } p_k^i. \tag{16}$$

The remainder in order to apply the random measurement matrix Kalman filtering is just to calculate:

$$R_{\tilde{\omega}_k} = R_{\omega_k} + E(\tilde{H}_k E(\mathbf{x}_k \mathbf{x}_k^T) \tilde{H}_k^T) = R_{\omega_k} + \sum_{i=1}^{l} p_k^i (H_k^i - \bar{H}_k) E(\mathbf{x}_k \mathbf{x}_k^T)(H_k^i - \bar{H}_k)^T \tag{17}$$

Substituting (15) and (17) into Theorem 1 can immediately obtain the random measurement matrix Kalman filtering of model (1), (13). In the following, two specific examples of the uncertain observations of the above model (1), (13) are given.

**Example 1.**

In the classical Kalman filtering problem, the observation is always assumed to contain the signal to be estimated. However, in practice, certain observation may contain noise alone, and the estimator cannot know this happens, only the probability of occurrence of such cases being available to the estimator. Nahi[4] derived the optimal recursive estimator with uncertain observation, but it is easy to see his result is a special case of ours except some notation difference.

Consider such a discrete dynamic process $\mathbf{x}_k, k = 0, 1, \cdots$ is defined by

$$\mathbf{x}_{k+1} = F_k \mathbf{x}_k + \nu_k, \tag{18}$$

where $F_k$ is a non-random matrix of appropriate dimension and $\nu_k$ is a noise sequence satisfying

$$E(\nu_k) = 0 \tag{19}$$

$$E(\nu_k \nu_l^T) = R_{\nu_k} \delta(k - l). \tag{20}$$

$\delta(\cdot)$ is the Kronecker delta function. The initial state $\mathbf{x}_0$ is assumed to be a random vector with a known mean $\mu_0$ and a known covariance matrix $P_0$.

The observation is given by

$$\begin{aligned}\mathbf{y}_k &= h_k \mathbf{x}_k + \omega_k, \text{ with probability } p(k) \\ &= \omega_k, \quad \text{ with probability } 1 - p(k),\end{aligned} \tag{21}$$

where $h_k$ is also an non-random matrix, $\omega_k$ is the observation noise satisfying

$$E(\omega_k) = 0 \tag{22}$$
$$E(\omega_k \omega_l^T) = R_{\omega_k} \delta(k-l). \tag{23}$$

$p(k)$ is the probability that the $kth$ observation contains the signal $\mathbf{x}_k$. Hence, the above observation can be described equivalently by

$$\mathbf{y}_k = H_k \mathbf{x}_k + \omega_k \tag{24}$$

where the observation matrix $H_k$ is a binary-valued random matrix, with

$$Pr\{H_k = h_k\} = p(k) \tag{25}$$
$$Pr\{H_k = \mathbf{0}\} = 1 - p(k) \tag{26}$$

Due to (8),

$$\bar{H}_k = p(k) \ h_k \tag{27}$$

$$\begin{aligned} Pr\{\tilde{H}_k = (1-p(k)) \ h_k\} = p(k) \\ Pr\{\tilde{H}_k = -p(k) \ h_k\} = 1 - p(k) \end{aligned} \tag{28}$$

In the uncertain observation case, the state transition matrix is still a constant one, but the measurement matrix is random, by (27) and (28), the covariance of the process and observation noise can be written as follows:

$$R_{\tilde{\nu}_k} = R_{\nu_k} \tag{29}$$

and

$$R_{\tilde{\omega}_k} = R_{\omega_k} + E(\tilde{H}_k E(\mathbf{x}_k \mathbf{x}_k^T) \tilde{H}_k^T) = R_{\omega_k} + (1-p(k))p(k) h_k E(\mathbf{x}_k \mathbf{x}_k^T) h_k^T. \tag{30}$$

Thus, the random measurement matrix Kalman Filtering in this special case is given by:

$$\begin{aligned} x_{k+1|k+1} &= x_{k+1|k} + K_{k+1}(y_{k+1} - p(k+1) h_{k+1} x_{k+1|k}) \\ x_{k+1|k} &= F_k x_{k|k} \\ K_{k+1} &= p(k+1) P_{k+1|k} h_{k+1}^T ( \ p(k+1)^2 h_{k+1} P_{k+1|k} h_{k+1}^T + R_{\tilde{\omega}_k})^+ \\ P_{k+1|k} &= F_k P_k F_k^T + R_{\nu_k} \\ P_{k+1} &= (I - p(k+1) K_{k+1} h_{k+1}) P_{k+1|k} \\ R_{\tilde{\omega}_k} &= R_{\omega_k} + (1-p(k))p(k) h_k E(x_k x_k^T) h_k^T \\ E(x_{k+1} x_{k+1}^T) &= F_k E(x_k x_k^T) F_k^T + R_{\nu_k} \\ x_{0|0} &= E(x_0), P_0 = Var(x_0), E(x_0 x_0^T) = E(x_0) E(x_0^T) + P_0. \end{aligned}$$

Compared the above formulas with Nahi's result [4], it is easy to see his result is a special case of ours except some notation difference.

**Example 2.**

We assume $\mathbf{y}_k$ has at least two elements and partition $\mathbf{y}_k$ into multiple parts, each part may contain noise alone. In the simplest case, suppose $\mathbf{y}_k$ is divided into two parts $\mathbf{y}_{k,1}, \mathbf{y}_{k,2}$. The observation equation can be given by:

$$\begin{pmatrix} \mathbf{y}_{k,1} \\ \mathbf{y}_{k,2} \end{pmatrix} = \begin{pmatrix} H_{k,1} \\ H_{k,2} \end{pmatrix} \mathbf{x}_k + \begin{pmatrix} \omega_{k,1} \\ \omega_{k,2} \end{pmatrix}, \tag{31}$$

where the observation matrix $H_{k,i}, i = 1, 2$ are independent of each other and two binary random matrices with

$$Pr\{H_{k,i} = h_{k,i}\} = p_i(k) \tag{32}$$
$$Pr\{H_{k,i} = \mathbf{0}\} = 1 - p_i(k) \tag{33}$$

Similarly as the derivation as before, we can obtain:

$$\bar{H}_k = \begin{pmatrix} p_1(k) h_{k,1} \\ p_2(k) h_{k,2} \end{pmatrix} \tag{34}$$

and the various samples of $\tilde{H}_k$ with their probabilities are given in the following table:

| Sam. of $\tilde{H}_k$ | $\begin{pmatrix} -p_1(k)h_{k,1} \\ -p_2(k)h_{k,2} \end{pmatrix}$ | $\begin{pmatrix} (1-p_1(k))h_{k,1} \\ (1-p_2(k))h_{k,2} \end{pmatrix}$ | $\begin{pmatrix} (1-p_1(k))h_{k,1} \\ -p_2(k)h_{k,2} \end{pmatrix}$ | $\begin{pmatrix} -p_1(k)h_{k,1} \\ (1-p_2(k))h_{k,2} \end{pmatrix}$ |
|---|---|---|---|---|
| $P_i$ | $(1-p_1(k))(1-p_2(k))$ | $p_1(k)p_2(k)$ | $p_1(k)(1-p_2(k))$ | $(1-p_1(k))p_2(k)$ |

Table 1: Samples of $\tilde{H}_k$ with their probabilities

Therefore,

$$\begin{aligned} R_{\tilde{\omega}_k} &= R_{\omega_k} + E(\tilde{H}_k E(\mathbf{x}_k \mathbf{x}_k^T) \tilde{H}_k^T) \\ &= R_{\omega_k} + \begin{pmatrix} (1-p_1(k))p_1(k) h_{k,1} E(\mathbf{x}_k \mathbf{x}_k^T) h_{k,1}^T & \mathbf{0} \\ \mathbf{0} & (1-p_2(k))p_2(k) h_{k,2} E(\mathbf{x}_k \mathbf{x}_k^T) h_{k,2}^T \end{pmatrix} \end{aligned} \tag{35}$$

Substituting (34) and (35) into the Kalman filtering in Theorem 1 can yield the optimum estimator straightforwardly for system (18), (31).

# 4  Application to Multiple-Model Dynamic Process

The multiple-model (MM) dynamic process were considered by many researchers (for example, see [7]-[11]). Although the possible models considered in those papers are quite general and can depend on the state, only suboptimal algorithms were proposed in the past a few decades. On the other hand, although some of the MM systems are not state-dependent and therefore more restrictive than the models considered in [8, 10], but these MM systems can be reduced to dynamic models with random transition matrix and thus the optimal real-time filter can be given directly according to the random transition matrix Kalman filtering proposed in Theorem 1.

Consider a system

$$\mathbf{x}_{k+1} = F_k^i \mathbf{x}_k + \nu_k \text{ with probability } p_k^i, \ i=1,2,\cdots,l. \tag{36}$$

$$\mathbf{y}_k = H_k \mathbf{x}_k + \omega_k \tag{37}$$

where $\{F_k^i\}$ and $\{\nu_k\}$ are independent sequences, and $H_k$ is non-random. We use random matrix $F_k$ to stand for the state transition matrix. The expectation of $F_k$ can be expressed as:

$$\bar{F}_k = \sum_{j=1}^{l} p_k^j F_k^j \tag{38}$$

$$\tilde{F}_k^i = F_k^i - \bar{F}_k, \text{ with probability } p_k^i \tag{39}$$

A necessary step for implemeting the random Kalman filtering is to calculate

$$R_{\tilde{\nu}_k} = R_{\nu_k} + E(\tilde{F}_k E(\mathbf{x}_k \mathbf{x}_k^T) \tilde{F}_k^T) = R_{\nu_k} + \sum_{i=1}^{l} p_k^i (F_k^i - \bar{F}_k) E(\mathbf{x}_k \mathbf{x}_k^T)(F_k^i - \bar{F}_k)^T \tag{40}$$

Thus, all the recursive formulas of random Kalman filtering can be given by:

$$\begin{aligned}
\mathbf{x}_{k+1|k+1} &= \mathbf{x}_{k+1|k} + K_{k+1}(y_{k+1} - H_{k+1}\mathbf{x}_{k+1|k}) \\
\mathbf{x}_{k+1|k} &= \bar{F}_k \mathbf{x}_{k|k} \\
K_{k+1} &= P_{k+1|k} H_{k+1}^T (H_{k+1} P_{k+1|k} H_{k+1}^T + R_{\omega_k})^+ \\
P_{k+1|k} &= \bar{F}_k P_k \bar{F}_k^T + R_{\tilde{\nu}_k} \\
P_{k+1} &= (I - K_{k+1} H_{k+1}) P_{k+1|k} \\
R_{\tilde{\nu}_k} &= R_{\nu_k} + \sum_{i=1}^{l} p_k^i (F_k^i - \bar{F}_k) E(\mathbf{x}_k \mathbf{x}_k^T)(F_k^i - \bar{F}_k)^T \\
E(\mathbf{x}_{k+1}\mathbf{x}_{k+1}^T) &= \bar{F}_k E(\mathbf{x}_k \mathbf{x}_k^T)\bar{F}_k^T + \sum_{i=1}^{l} p_k^i (F_k^i - \bar{F}_k) E(\mathbf{x}_k \mathbf{x}_k^T)(F_k^i - \bar{F}_k)^T + R_{\nu_k} \\
\mathbf{x}_{0|0} &= E\mathbf{x}_0, \ P_0 = Var(\mathbf{x}_0), \ E(\mathbf{x}_0 \mathbf{x}_0^T) = E\mathbf{x}_0 E\mathbf{x}_0^T + P_0,
\end{aligned}$$

# 5  Simulations

The simulations were done for a dynamic system with random parameter matrices modelled as an object movement with process noise and measurement noise on the plane. The two simulations show the specific applications of results in the last two sections.

**Simulation 1.**  We consider the model in example 1, and certain observations may contain noise alone, only the probability of occurrence available to the estimator. The object dynamics and measurement equations are given by,

$$F_k = \begin{pmatrix} cos(2\pi/300) & sin(2\pi/300) \\ -sin(2\pi/300) & cos(2\pi/300) \end{pmatrix} \tag{41}$$

$$H_k = \begin{pmatrix} 1 & 1 \\ 1 & -1 \end{pmatrix}, \text{ with probability } \gamma = 0.95 \tag{42}$$

$$= \mathbf{0}, \text{ with probability } 1 - \gamma = 0.05 \tag{43}$$

The initial state $\mathbf{x}_0 = (50, 0)$, $P_{0|0} = 0.5I$. The covariance of the noises are diagonal, given by $R_\nu = 2$, $R_\omega = 1$. It is easy to see that the target is a object that moves noisily at constant rotation speed $2\pi/300$/step in a circle with initial radius 50 about origin of the coordinate space. Using a tracking trajectory and Monte-Carlo method of 50 runs, we can evaluate tracking performance of an algorithm by comparing a tracking trajectory with the actual moving object (see Fig.1 below) and showing the second moment of the tracking error (see Fig. 2 below) given by

$$E_k^2 = \frac{1}{50} \sum_{i=1}^{50} ||\mathbf{x}_{k|k}^{(i)} - \mathbf{x}_k||^2$$

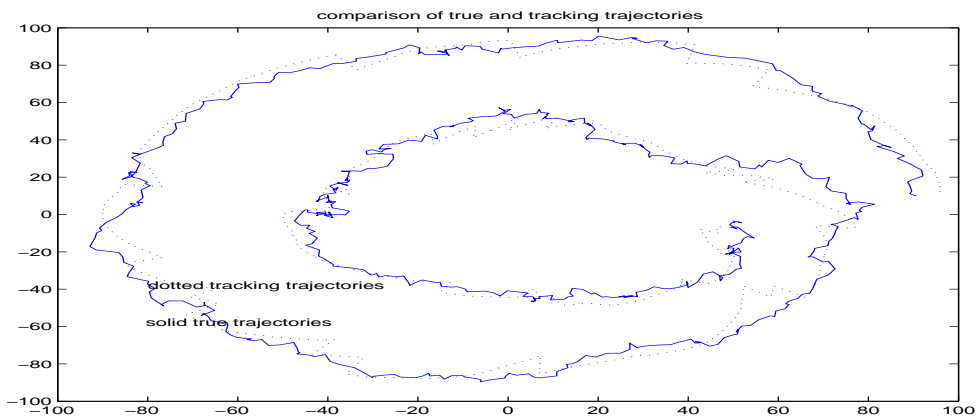

Figure 1: Tracking Trajectory of Random KAL with two measurement equations

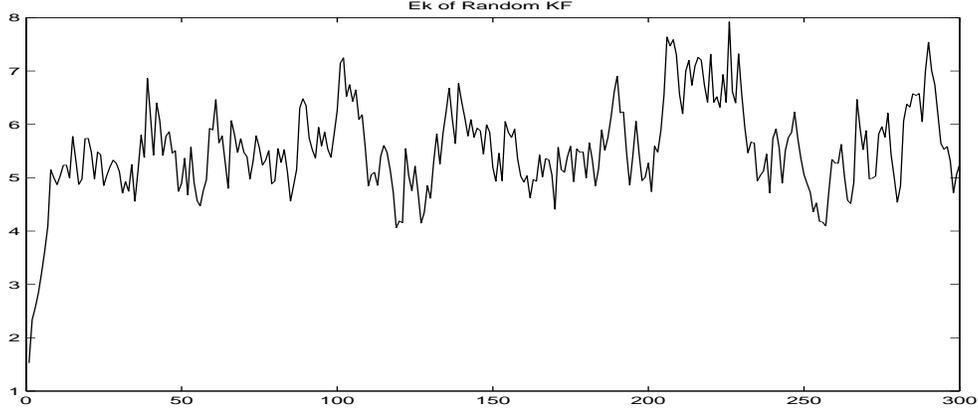

Figure 2: Estimated tracking error variance of Random KAL with two measurement equations

From Figs.1 and 2, the tracking performance of the random Kalman filtering looks acceptable, and the more simulations have done for the different probabilities of observation $\gamma$. The those simulation results show that the smaller $\gamma$ is, the bigger the tracking error is, which is consistent with intuitive expectation.

**Simulation 2.** In this simulation, there are three dynamic models , with the corresponding probabilities of occurrence available. The object dynamics and measurement equations are given by,

$$F_k = \begin{cases} \begin{pmatrix} cos(2\pi/300) & sin(2\pi/300) \\ -sin(2\pi/300) & cos(2\pi/300) \end{pmatrix} \text{ with probability } 0.1, \\ \begin{pmatrix} cos(2\pi/250) & sin(2\pi/250) \\ -sin(2\pi/250) & cos(2\pi/250) \end{pmatrix} \text{ with probability } 0.2, \\ \begin{pmatrix} cos(2\pi/100) & sin(2\pi/100) \\ -sin(2\pi/100) & cos(2\pi/100) \end{pmatrix} \text{ with probability } 0.7, \end{cases} \quad (44)$$

$$H_k = \begin{pmatrix} 1 & 1 \\ 1 & -1 \end{pmatrix}. \quad (45)$$

Obviously, for this system, the object at time $k$ moves at three different rotation speeds with the two corresponding probabilities, respectively, in a circle with radius $\|\mathbf{x}_k\|$ about origin of the coordinate space.

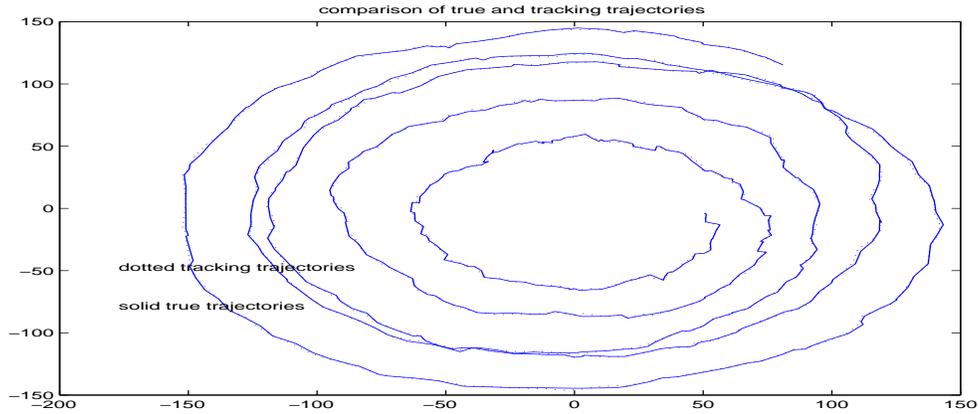

Figure 3: Tracking trajectory of random KAL with three dynamics equations

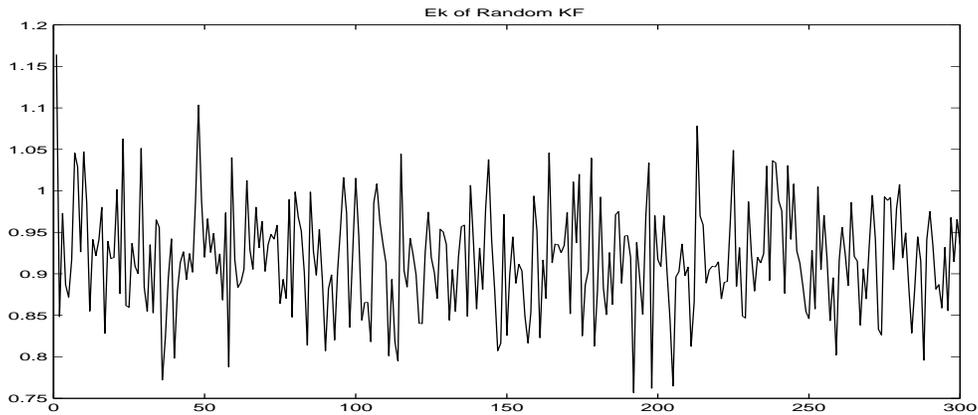

Figure 4: Estimated tracking error variance of random KAL with three dynamics equations

From Figs.3 and 4, it can be shown that the filter given in section 4 does work well.

## 6  Conclusion

In this paper, we have given rigorous analysis for the Linear Minimum Variance recursive state estimation of the linear discrete time dynamic system with random state transition and measurement matrices. Since such a system can be converted to a linear dynamic system with deterministic parameter matrices and state-dependent process and measurement noises. We have shown that under mild conditions, the converted system still satisfies the conditions of standard Kalman Filtering; therefore, the recursive state estimation of this system is still of the form of a modified Kalman filtering. More importantly, we found that this result can be applied to Kalman filtering with uncertain observations as well as randomly variant dynamic systems with multiple models. The simulation examples support our analysis for the applications of the random parameter matrices Kalman filtering.

# Appendix

In this appendix, we provide proofs for the various lemmas that are presented in the paper.

**Proof of Lemma 1.**
By the properties of conditional expectation, we have that

$$
\begin{aligned}
E(F\mathbf{x}\mathbf{x}^T F^T) &= E(E(F\mathbf{x}\mathbf{x}^T F^T | F)) \\
&= E(FE(\mathbf{x}\mathbf{x}^T | F)F^T) \\
&= E(FE(\mathbf{x}\mathbf{x}^T)F^T). \quad \text{(A.1)}
\end{aligned}
$$

**Q.E.D.**

**Proof of Lemma 2.**
$(a)$ : By the assumptions on the model (1) and (2), and notations in (11), it is obvious.

$(b1)$ : Since $\{F_k, \nu_k, k = 0, 1, 2, \ldots\}$ is independent of $\mathbf{x}_0$,

$$
\begin{aligned}
E(\mathbf{x}_0 \tilde{\nu}_k^T) &= E\{\mathbf{x}_0 (\nu_k + \tilde{F}_k \mathbf{x}_k)^T\} \\
&= E(\mathbf{x}_0 \mathbf{x}_k^T \tilde{F}_k^T) \\
&= E\{\mathbf{x}_0 (F_{k-1}\mathbf{x}_{k-1} + \nu_{k-1})^T \tilde{F}_k^T\} \\
&= E(\mathbf{x}_0 \mathbf{x}_{k-1}^T F_{k-1}^T \tilde{F}_k^T) \\
&= E\{\mathbf{x}_0 (F_{k-2}\mathbf{x}_{k-2} + \nu_{k-2})^T F_{k-1}^T \tilde{F}_k^T\} \\
&= E(\mathbf{x}_0 \mathbf{x}_{k-2}^T F_{k-2}^T F_{k-1}^T \tilde{F}_k^T) \\
&\quad \cdots \\
&= E(\mathbf{x}_0 \mathbf{x}_0^T F_0^T F_1^T \cdots F_{k-2}^T F_{k-1}^T \tilde{F}_k^T) \\
&= E(\mathbf{x}_0 \mathbf{x}_0^T) E(F_0^T) E(F_1^T) \cdots E(F_{k-2}^T) E(F_{k-1}^T) E(\tilde{F}_k^T) \\
&= 0.
\end{aligned}
$$

$(b2)$ : Similarly, $E(\mathbf{x}_0 \tilde{\omega}_k^T) = 0$.

$(c1)$ : Without loss of generality, we consider the case of $k > l$ only.

$$E(\tilde{\nu}_k \tilde{\nu}_l^T) = E(\nu_k \nu_l^T + \nu_k \mathbf{x}_l^T \tilde{F}_l^T + \tilde{F}_k \mathbf{x}_k \nu_l^T + \tilde{F}_k \mathbf{x}_k \mathbf{x}_l^T \tilde{F}_l^T)$$

For $\mathbf{x}_l$ is linearly dependent on $F_{l-1} F_{l-2} \cdots F_1 F_0 \mathbf{x}_0, \nu_{l-1}, F_{l-1}F_{l-2} \cdots F_{l-i+1}\nu_{l-i}, i = 2, 3, \cdots, l$, and $\{F_k, \nu_k, k = 0, 1, 2, \ldots\}$ is independent of $\mathbf{x}_0$,

$$
\begin{aligned}
E(\nu_k x_l^T \tilde{F}_l^T) &= 0 \\
E(\tilde{F}_k \mathbf{x}_k \nu_l^T) &= E\{\tilde{F}_k(F_{k-1}\mathbf{x}_{k-1} + \nu_{k-1})\nu_l^T\} \\
&= E(\tilde{F}_k F_{k-1}\mathbf{x}_{k-1}\nu_l^T) \\
&= E\{\tilde{F}_k F_{k-1}(F_{k-2}\mathbf{x}_{k-2} + \nu_{k-2})\nu_l^T\} \\
&= E(\tilde{F}_k F_{k-1} F_{k-2}\mathbf{x}_{k-2}\nu_l^T) + E(\tilde{F}_k F_{k-1}\nu_{k-2}\nu_l^T) \\
&\cdots \\
&= E(\tilde{F}_k F_{k-1} \cdots F_l \mathbf{x}_l \nu_l^T) + E(\tilde{F}_k F_{k-1} \cdots F_{l+1}\nu_l \nu_l^T) \\
&= E(\tilde{F}_k) E(F_{k-1}) \cdots E(F_{l+1}) E(\nu_l \nu_l^T) \\
&= 0. \quad\quad\quad\quad\quad\quad\quad\quad\quad\quad\quad\quad\quad\quad\quad\quad\quad\quad\quad\quad\quad\quad\quad\quad\quad\quad (A.2)
\end{aligned}
$$

Noting that $\mathbf{x}_l$ and $\{F_k, k = 0, 1, 2, \cdots\}$ are independent, we have

$$
\begin{aligned}
E(\tilde{F}_k \mathbf{x}_k \mathbf{x}_l^T \tilde{F}_l^T) &= E\{\tilde{F}_k(F_{k-1}\mathbf{x}_{k-1} + \nu_{k-1})\mathbf{x}_l^T \tilde{F}_l^T)\} \\
&= E(\tilde{F}_k F_{k-1}\mathbf{x}_{k-1}\mathbf{x}_l^T \tilde{F}_l^T) \\
&= E(\tilde{F}_k F_{k-1} F_{k-2}\mathbf{x}_{k-2}\mathbf{x}_l^T \tilde{F}_l^T + \tilde{F}_k F_{k-1}\nu_{k-2}\mathbf{x}_l^T \tilde{F}_l^T) \\
&\cdots \\
&= E(\tilde{F}_k F_{k-1} F_{k-2}\cdots F_{l+1} F_l \mathbf{x}_l \mathbf{x}_l^T \tilde{F}_l^T) + E(\tilde{F}_k F_{k-1} F_{k-2}\cdots F_{l+1}\nu_l \mathbf{x}_l^T \tilde{F}_l^T) \\
&= E(\tilde{F}_k F_{k-1} F_{k-2}\cdots F_{l+1} F_l \mathbf{x}_l \mathbf{x}_l^T \tilde{F}_l^T) \\
&= E(\tilde{F}_k F_{k-1} F_{k-2}\cdots F_{l+1} \bar{F}_l \mathbf{x}_l \mathbf{x}_l^T \tilde{F}_l^T) + E(\tilde{F}_k F_{k-1} F_{k-2}\cdots F_{l+1} \tilde{F}_l \mathbf{x}_l \mathbf{x}_l^T \tilde{F}_l^T) \\
&= E(E(\tilde{F}_k F_{k-1} F_{k-2}\cdots F_{l+1} \bar{F}_l) E(\mathbf{x}_l \mathbf{x}_l^T) \tilde{F}_l^T) + E(E(\tilde{F}_k F_{k-1} F_{k-2}\cdots F_{l+1}) \tilde{F}_l E(\mathbf{x}_l \mathbf{x}_l^T) \tilde{F}_l^T) \\
&= 0. \quad\quad\quad\quad\quad\quad\quad\quad\quad\quad\quad\quad\quad\quad\quad\quad\quad\quad\quad\quad\quad\quad\quad\quad\quad (A.3)
\end{aligned}
$$

Hence, $E(\tilde{\nu}_k \tilde{\nu}_l^T) = 0$.

(c2) : Also consider the case of $k > l$ only. Since $\mathbf{x}_l$ is linearly dependent on $F_{l-1} F_{l-2}\cdots F_1 F_0 \mathbf{x}_0$, $\nu_{l-1}$, $F_{l-1} F_{l-2}\cdots F_{l-i+1}\nu_{l-i}$, $i = 2, 3, \cdots, l$, $\{F_k, H_k, \nu_k, \omega_k, k = 0, 1, 2, \ldots\}$ is independent of $\mathbf{x}_0$, and $\{F_k, H_k, k = 0, 1, 2, \ldots\}$ is independent of $\mathbf{x}_l$. Moreover,

$$
\begin{aligned}
E(\tilde{\omega}_k \tilde{\omega}_l^T) &= E(\omega_k \omega_l^T + \omega_k \mathbf{x}_l^T \tilde{H}_l^T + \tilde{H}_k \mathbf{x}_k \omega_l^T + \tilde{H}_k \mathbf{x}_k \mathbf{x}_l^T \tilde{H}_l^T) \\
&= E(\tilde{H}_k \mathbf{x}_k \mathbf{x}_l^T \tilde{H}_l^T) \\
&= E\{\tilde{H}_k(F_{k-1}\mathbf{x}_{k-1} + \nu_{k-1})\mathbf{x}_l^T \tilde{H}_l^T\} \\
&= E(\tilde{H}_k F_{k-1}\mathbf{x}_{k-1} x_l^T \tilde{H}_l^T) \\
&= E(\tilde{H}_k F_{k-1} F_{k-2}\mathbf{x}_{k-2}\mathbf{x}_l^T \tilde{H}_l^T + \tilde{H}_k F_{k-1}\nu_{k-2}\mathbf{x}_l^T \tilde{H}_l^T) \\
&\cdots \\
&= E(\tilde{H}_k F_{k-1} F_{k-2}\cdots F_{l+1} F_l \mathbf{x}_l \mathbf{x}_l^T \tilde{H}_l^T) + E(\tilde{H}_k F_{k-1} F_{k-2}\cdots F_{l+1}\nu_l \mathbf{x}_l^T \tilde{H}_l^T)
\end{aligned}
$$

$$= 0. \tag{A.4}$$

$(c3)$ :
$$E(\tilde{\nu}_k \tilde{\omega}_l^T) = E(\nu_k \omega_l^T + \nu_k \mathbf{x}_l^T \tilde{H}_l^T + \tilde{F}_k \mathbf{x}_k \omega_l^T + \tilde{F}_k \mathbf{x}_k \mathbf{x}_l^T \tilde{H}_l^T).$$

When $k \geq l$, as the derivation in (A.4), we can obtain,

$$\begin{aligned} E(\tilde{\nu}_k \tilde{\omega}_l^T) &= E(\tilde{F}_k \mathbf{x}_k \mathbf{x}_l^T \tilde{H}_l^T) \\ &= E\{\tilde{F}_k (F_{k-1} \mathbf{x}_{k-1} + \nu_{k-1}) \mathbf{x}_l^T \tilde{H}_l^T\} \\ &= E(\tilde{F}_k F_{k-1} \mathbf{x}_{k-1} \mathbf{x}_l^T \tilde{H}_l^T) \\ &= E(\tilde{F}_k F_{k-1} F_{k-2} \mathbf{x}_{k-2} x_l^T \tilde{H}_l^T) + E(\tilde{F}_k F_{k-1} \nu_{k-2} \mathbf{x}_l^T \tilde{H}_l^T) \\ &\cdots \\ &= E(\tilde{F}_k F_{k-1} F_{k-2} \cdots F_{l+1} F_l \mathbf{x}_l \mathbf{x}_l^T \tilde{H}_l^T + \tilde{F}_k F_{k-1} F_{k-2} \cdots F_{l+1} \nu_l \mathbf{x}_l^T \tilde{H}_l^T) \\ &= 0. \end{aligned}$$

When $k < l$,, as derivation in (A.2) and (A.3),

$$\begin{aligned} E(\nu_k \mathbf{x}_l^T \tilde{H}_l^T) &= E\{\nu_k (F_{l-1} \mathbf{x}_{l-1} + \nu_{l-1})^T \tilde{H}_l^T\} \\ &= E(\nu_k \mathbf{x}_{l-1}^T F_{l-1}^T \tilde{H}_l^T) \\ &= E(\nu_k \mathbf{x}_{l-2}^T F_{l-2}^T F_{l-1}^T \tilde{H}_l^T) + E(\nu_k \nu_{l-2}^T F_{l-1}^T \tilde{H}_l^T) \\ &\cdots \\ &= E(\nu_k \mathbf{x}_k^T F_k^T F_{k+1}^T \cdots F_{l-1}^T \tilde{H}_l^T) + E(\nu_k \nu_k^T F_{k+1}^T \cdots F_{l-1}^T \tilde{H}_l^T) \\ &= 0, \\ E(\tilde{F}_k \mathbf{x}_k \mathbf{x}_l^T \tilde{H}_l^T) &= E\{\tilde{F}_k \mathbf{x}_k (F_{l-1} \mathbf{x}_{l-1} + \nu_{l-1})^T \tilde{H}_l^T\} \\ &= E(\tilde{F}_k \mathbf{x}_k \mathbf{x}_{l-1}^T F_{l-1}^T \tilde{H}_l^T) \\ &= E\{\tilde{F}_k \mathbf{x}_k (F_{l-2} \mathbf{x}_{l-2} + \nu_{l-2})^T F_{l-1}^T \tilde{H}_l^T\} \\ &\cdots \\ &= E\{\tilde{F}_k \mathbf{x}_k (F_k \mathbf{x}_k + \nu_k)^T F_{k+1}^T \cdots F_{l-1}^T \tilde{H}_l^T\} \\ &= E(\tilde{F}_k \mathbf{x}_k \mathbf{x}_k^T F_k^T F_{k+1}^T \cdots F_{l-1}^T \tilde{H}_l^T) \\ &= E(\tilde{F}_k E(\mathbf{x}_k \mathbf{x}_k^T) F_k^T E(F_{k+1}^T \cdots F_{l-1}^T \tilde{H}_l^T)) \\ &= 0. \end{aligned}$$

Thus, $E(\tilde{\nu}_k \tilde{\omega}_l^T) = 0$.

$(d)$ :
$$\begin{aligned} E(\tilde{\nu}_k \tilde{\nu}_k^T) &= E(\nu_k \nu_k^T + \nu_k \mathbf{x}_k^T \tilde{F}_k^T + \tilde{F}_k \mathbf{x}_k \nu_k^T + \tilde{F}_k \mathbf{x}_k \mathbf{x}_k^T \tilde{F}_k^T) \\ &= R_{\nu_k} + E(\tilde{F}_k \mathbf{x}_k \mathbf{x}_k^T \tilde{F}_k^T) \\ &= R_{\nu_k} + E(\tilde{F}_k E(\mathbf{x}_k \mathbf{x}_k^T) \tilde{F}_k^T) \end{aligned}$$

Similarly, $E(\tilde{\omega}_k \tilde{\omega}_k^T) = R_{\omega_k} + E(\tilde{H}_k E(\mathbf{x}_k \mathbf{x}_k^T) \tilde{H}_k^T)$.

**Q.E.D.**